\documentclass[article]{IEEEtran}
\ifCLASSINFOpdf
\else
\fi
\makeatletter
\def\ps@headings{%
\def\@oddhead{\mbox{}\scriptsize\rightmark \hfil \thepage}%
\def\@evenhead{\scriptsize\thepage \hfil \leftmark\mbox{}}%
\def\@oddfoot{}%
\def\@evenfoot{}}
\makeatother 

\IEEEoverridecommandlockouts
\usepackage{color}
\usepackage{amsfonts}
\usepackage[dvips]{graphicx}
\usepackage{caption}
\usepackage{subfigure}
\usepackage{times}
\usepackage{cite}
\usepackage{lettrine}
\usepackage{amsmath}
\usepackage{amsmath}
\allowdisplaybreaks[4]
\usepackage{array}
\usepackage{amssymb}

\usepackage{stfloats}
\usepackage{graphicx}
\usepackage{footnote}
\usepackage{booktabs}
\usepackage{array}
\usepackage{algorithmic}
\usepackage{algorithm}
\usepackage{subeqnarray}
\usepackage{cases}
\usepackage{threeparttable}
\usepackage{color}

\newtheorem{lemma}{\underline{Lemma}}[section]

\newtheorem{proposition}{\underline{Proposition}}[section]

\begin{document}
\bibliographystyle{IEEEtran}

\title{Energy-Efficient SWIPT in IoT Distributed Antenna Systems}
\IEEEoverridecommandlockouts

\author{\IEEEauthorblockN{Yuwen Huang, Mengyu Liu, and Yuan Liu,~\IEEEmembership{Member,~IEEE} }

\thanks{Manuscript received October 3, 2017; accept January 8, 2018. This paper was supported in part by the Natural Science Foundation of China under Grant 61401159, Grant 61771203 and Grant U1701265, and in part by the Pearl River Science and Technology Nova Program of Guangzhou under Grant 201710010111. \emph{(Corresponding author: Y. Liu)}

The authors are with School of Electronic and Information Engineering, South China University of Technology, Guangzhou 510641, China (email: eehyw@mail.scut.edu.cn, liu.mengyu@mail.scut.edu.cn, eeyliu@scut.edu.cn). }
}
\maketitle

\vspace{-1.5cm}
\maketitle
 \begin{abstract}
The rapid growth of Internet of Things (IoT) dramatically increases power consumption of wireless devices. Simultaneous wireless information and power transfer (SWIPT) is a promising solution for sustainable operation of IoT devices. In this paper, we study energy efficiency (EE) in SWIPT-based distributed antenna system (DAS), where  power splitting (PS) is applied at IoT devices to coordinate the energy harvesting (EH) and information decoding (ID) processes by varying transmit power of distributed antenna (DA) ports and PS ratios of IoT devices. In the case of single IoT device, we find the optimal closed-form solution by deriving some useful properties based on Karush-Kuhn-Tucker (KKT) conditions and the solution is no need for numerical iterations. For the case  of multiple IoT devices, we propose an efficient suboptimal algorithm to solve the EE maximization problem. Simulation results show that the proposed schemes achieve better EE performance compared with other benchmark schemes in both single and multiple IoT devices cases.
 \end{abstract}


 \section{Introduction}

 Next generation communication systems are expected to support billions of wireless devices due to the advancement of Internet of things (IoT), which leads to the growing energy consumption and has triggered a dramatic increase of research in energy consumption of wireless communications. Due to the sharply growing energy costs and the drastic greenhouse gas increase, green communication or energy-efficient wireless communication, has drawn a wide attraction recently. Therefore, pursing higher data transmission rate as well as lowering energy consumption is the trend toward future IoT networks. The energy efficiency (EE) is defined as the sum-rate divided by the total power consumption and is measured by bit/Hz/Joule. So far, a large number of technologies/methods  have been studied for improving the EE performance in a variety of wireless communication systems \cite{Huang2014,Huang2016,Huang2017,Huang2016a}.

Recently, distributed antenna system (DAS)  has gained its popularity in the next generation communication systems due to its advantage in increasing both EE and spectral efficiency (SE) by expanding system's coverage and improving the sum achievable rate \cite{Lee2012,Choi2007,Hasegawa2003,Lee2013}. In conventional cellular systems, the antennas are co-located at the base station and in charge of baseband signal processing as well as radio frequency (RF) operations. Distinguished from a conventional antenna system (CAS) with centralized antennas and base station at the center location, a promising technique, i.e., DAS, is introduced for next generation cellular systems by splitting the functionalities of the base stations into a central processor (CP) and distributed antenna (DA) ports. In DAS, DA ports are separate geographically in the cell with independent power supply and connected to the CP via high capacity optical fibers or cables. Especially, the CP performs computationally intensive baseband signal processing and the DA ports are in charge of all RF operations such as analog filtering and amplifying. As a result, the overall performance of the system can be enhanced by narrowing the access distances between the devices and the DA ports. In particular, the SE analysis in terms of the downlink capacity of DAS under a single device environment was studied in \cite{Lee2012} for the cases with and without perfect channel state information (CSI) at the transmitters according to the information theoretic view. Note that DAS gains its popularity as a highly promising candidate for the 5G mobile communication systems and has been applied to many advanced technologies such as the cloud radio access network (CRAN) \cite{Peng2015}.

To meet the concept of energy-efficient wireless communication, EE optimization in DAS has been widely studied in the literature \cite{He2013,Li2016,Zhang2010,Chen2012,He2014,Kim2015,Kim2012}. The authors in \cite{He2013} solved the EE maximization problem with proportional fairness consideration. An energy-efficient scheme of joint antenna, subcarrier, and power allocation was studied in \cite{Li2016}. An energy-efficient DAS layout with multiple sectored antennas was proposed in \cite{Zhang2010}. The optimal energy-efficient power allocation problem was studied in generalized DAS \cite{Chen2012}. The authors in \cite{He2014} developed an energy-efficient resource allocation scheme with proportional fairness for downlink orthogonal frequency-division multiplexing access (OFDMA) DAS. In \cite{Kim2015}, the authors considered an optimal power allocation scheme in DAS and provided a simplified scheme where the user is served by a single DA port with the best channel gain. Compared with the optimal algorithm, this scheme performs little EE loss with remarkably reductions in system's overhead. However, for SE maximization problem in DAS, serving a user with fewer DA ports or less transmission power achieves worse SE than transmitting full power at all active DA ports \cite{Kim2012}.

Energy harvesting (EH) has been introduced as a promising solution to prolong lifetime of the energy-constrained IoT devices. The same radio-frequency (RF) signals can be used for simultaneous wireless information and power transfer (SWIPT). Two practical receiver designs were proposed in \cite{Liu2013,Zhang2013}, namely ``time switching'' (TS) and ``power splitting'' (PS). In particular, the TS receiver switches between decoding information and harvesting energy for the received signals, while the PS receiver splits the received signals into two streams for information decoding (ID) and EH with a PS ratio. In \cite{Jiang2017} and \cite{Xiong2017}, the authors considered a non-linear EH model for EH. Furthermore, channel statistics in SWIPT was studied in \cite{Mishra2017}. A variety of resource allocation schemes were  studied for SWIPT systems \cite{Ng2013,Liu2016,Liu2016a,Xiong2015,Zhang2016,Zhang2016a}.

A challenge of applying SWIPT in IoT networks is the fast decay of energy transfer efficiency over the transmission distance. However, this problem can be alleviated in DAS due to the short transmitter-receiver distances. As a result, there is performance potential by integrating DAS into SWIPT. The combination of these two technologies is in accordance with the importance of energy-efficient IoT network. There are a handful of works studying SWIPT-based DAS. For instance, the authors in \cite{Ng2015} focused secure SWIPT in DAS for transmit power minimization. Joint wireless information and energy transfer was investigated in massive DAS \cite{Yuan2015}. However, to our best knowledge, there is no work considering EE in SWIPT-based DAS.

The main contributions of this paper are summarized as follows:
\begin{itemize}
	\item We study the EE optimization problem in SWIPT-based DAS, where PS is applied at the IoT devices to coordinate EH and ID processes. Our goal is to maximize the system's EE while satisfying the minimum harvested energy requirements of the IoT devices and individual power budgets of the DA ports.

\item For the case of single IoT device, the EE problem is a non-convex problem. Unlike the traditional methods of solving the fractional programming problems, we find the globally optimal solution by analyzing the Karush-Kuhn-Tucker (KKT) conditions, which has the closed-form and is no need for any numerical search or iteration.

\item The EE maxmization problem of multiple IoT devices case is also non-convex. We propose a two-step suboptimal algorithm. Specifically, at the first step, assuming the PS ratios of the IoT devices are given, we optimize the transmit power of the DA ports using the block coordinate descent (BCD) method. At the second step, we find the optimal PS ratio of each IoT device for given transmit power of the DA ports.
\end{itemize}
	
The rest of this paper is organized as follows. Section II presents optimal power allocation policy for EE maximization in DAS with a single IoT device. Section III details the proposed suboptimal algorithm for the case of multiple IoT devices. Sections IV presents simulation results and discussions. Finally, Section V concludes the paper.

\section{Single IoT Device Case}
\subsection{System Model and Problem Formulation}
\begin{figure}[t]
\begin{centering}
\includegraphics[scale=0.6]{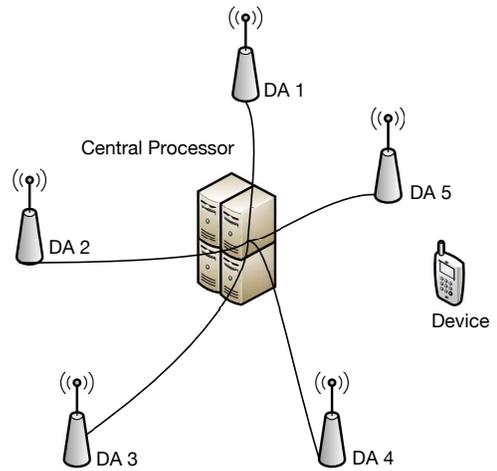}
\vspace{-0.1cm}
 \caption{ An example of system model of DAS with five DA ports. }\label{fig:system}
\end{centering}
\vspace{-0.1cm}
\end{figure}
\begin{figure}[t]
\begin{centering}
\includegraphics[scale=0.6]{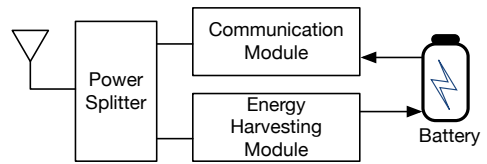}
\vspace{-0.1cm}
 \caption{ The receiver circuit of the SWIPT based IoT devices. }\label{fig:receiver}
\end{centering}
\vspace{-0.1cm}
\end{figure}

We consider a downlink single cell DAS with $N$ DA ports, a CP and a device, as shown in Fig. \ref{fig:system}, where each DA port is equipped with a single antenna. Here we plot the receiver circuit of the SWIPT based IoT devices, as shown in Fig. \ref{fig:receiver}. The device has a power splitter so that its received signal power is split into two parts, with $0\leq\alpha\leq1$ for ID and the rest $1-\alpha$ for EH. Let $p_i$ and $h_{i}$ denote the transmit power of DA port $i$ and the channel power gain from DA port $i$ to the device, respectively. In addition, $\sigma^2$ represents the independent and identically distributed circularly symmetric complex Gaussian noise. In our system model, we assume that the perfect CSI is available. This assumption is also reasonable for low-end SWIPT-based IoT devices. This is because that the low-end IoT devices are general information nodes which must have the basic communication function to transmit/decode information signals. As a result, the IoT devices can transmit/decode pilot or reference signals, which is much easier since pilot or reference signal are known signals. Similar to wireless communication systems, there are many ways to obtain the CSI for the SWIPT-based IoT devices, like training at either uplink or downlink using pilot signals \cite{Chen2014,Yang2014,Zhou2015}, or simple 1-bit feedback \cite{Xu2014}. To sum up, it is possible to obtain CSI for IoT devices as they are information nodes with communication module.

Thus the achievable rate for the device is expressed as
\begin{align}\label{eqn:r01}
R=\ln\left(1+\frac{\alpha\sum_{i=1}^N h_{i}p_{i}}{\sigma^{2}}\right).
\end{align}
Because in DAS, the DA ports are separate geographically in the cell with independent power supply and connected to the CP via optical cable. To be more realistic, we suppose that each DA port works independently and does not share power with other DA ports. Every DA port has its own transmit power constraint, which is expressed as $p_{i}\le \bar{P}_i$ for all $i$. The EE for the DAS is defined as
\begin{align}\label{eqn:r1}
\eta_{1}=\frac{R}{\sum_{i=1}^Np_{i}+p_{c}},
\end{align}
where $p_{c}$ denotes  the circuit power consumption. We can see that the power consumption is divided into two parts: the consumption of power amplifiers in each DA port and the circuit parts' consumption power $p_{c}$ which includes the power used to run the digital signal processors, mixers and so on. Here we consider the linear EH model with linear energy conversion efficiency $\zeta$. In practice, $\zeta$ is non-linear in general. However, as shown in \cite{Di2017}, the linear EH model is still meaningful for the following reasons. First, the linear EH model is more trackable and the non-linear one shows piecewise linearity in the relative low and high input power cases. Second, due to signal attenuation, the EH devices have a high possibility to work in the low input power case, which can be approximated as a linear model. Therefore, in this paper we consider the linear EH model for the purpose of more tractable analysis. Denoting $0<\zeta\leq1$ as the energy conversion efficiency, the harvested energy in Joule at the energy receiver can be written as
\begin{align}\label{eqn:r2}
E=\zeta(1-\alpha)\sum_{i=1}^{N}h_{i}p_{i}.
\end{align}

Our objective is to maximize the system's EE by varying the transmit power of DA ports and the PS ratio of the device, subjected to the minimum harvested power requirement $\bar{E}$ for the device and the maximum transmit power constraint $\bar{P}_i$ for each DA port. Thus the problem is formulated as
\begin{eqnarray}\label{eqn:r3}
 {\rm (P1):}\nonumber
 \max_{{\alpha},\{p_{i}\}}&&\eta_{1}  \\
{\rm s.t.}&&E\ge \bar{E},\label{eqn:r113}\\
  && 0\leq \alpha\leq 1,\label{eqn:r69}\\
  && 0\leq p_{i}\leq \bar{P}_i,i=1,\cdots,N.\label{eqn:r70}
 \end{eqnarray}
\subsection{Proposed Optimal Solution}
Problem (P1) is a non-convex problem, since the objective function of (P1) $\eta_1$ is a non-concave function. In this subsection, we propose an optimal solution to solve Problem (P1).

We analyze the KKT conditions for Problem (P1) first. The Lagrangian function for Problem (P1) is written as
\begin{align}\label{eqn:r16}
&L_1(\{p_{i}\},\alpha,\mu,\{\lambda_i\},\{\upsilon_i\})=\nonumber\\&\frac{\ln\left(1+\frac{\alpha}{\sigma^{2}}\sum_{i=1}^N h_{i}p_{i}\right)\nonumber}{\sum_{i=1}^Np_{i}+p_{c}}+\sum_{i=1}^{N}\lambda_{i}p_{i}  +\sum_{i=1}^{N}\upsilon_{i}\left(\bar{P}_i-p_{i}\right)\nonumber\\&+\mu\left[\zeta(1-\alpha)\sum_{i=1}^{N}h_{i}p_{i}-\bar{E}\right],
\end{align}
where $\{\lambda_{i}\}$ and $\{\upsilon_{i}\}$ are the Lagrangian multipliers with respect to the constraints $p_{i}\ge 0$ and $p_{i}\le \bar{P}_i$, for all $i$. In addition, $\mu$ is associated with the constraint \eqref{eqn:r113}. The dual function of Problem (P1) is given by
\begin{equation}\label{eqn:r111}
g_1(\mu,\{\lambda_i\},\{\upsilon_i\})=\max_{ 0\leq \alpha\leq 1}L_1(\{p_{i}\},\alpha,\mu,\{\lambda_i\},\{\upsilon_i\}).
\end{equation}
According to the KKT conditions, the optimal value $\{\lambda^{*}_{i},\upsilon^{*}_{i},p_{i}^{*},\alpha^*,\mu^*\}$ for $i=1,\cdots,N$ should satisfy the following conditions:
\begin{eqnarray}\label{eqn:r17}
&&\frac{\partial L_1}{\partial p_{i}}=f_{i}(p_{1}^{*},\cdots,p_{N}^{*},\alpha^*)+\lambda^{*}_{i}-\upsilon^{*}_{i}=0, \label{eqn:r47} \\
&&\frac{\partial L_1}{\partial \alpha}=\sum_{i=1}^{N}h_{i}p_{i} \left( T(p_{1}^{*},\cdots,p_{n}^{*},\alpha^*)-\zeta\mu^*\right)=0, \label{eqn:r48} \\
&&\lambda^{*}_{i}p^{*}_{i}=\upsilon^{*}_{i}\left(\bar{P}_i-p^{*}_{i}\right)=0, \label{eqn:r18} i=1, \cdots, N,\\
&&\mu^*\left(\zeta(1-\alpha^*)\sum_{i=1}^{N}h_{i}p^*_{i}- \bar{E}\right)=0, \label{eqn:r51}\\
&& 0\le p_{i}^* \le \bar{P}_i, i=1, \cdots, N,\nonumber\\
&&\lambda^{*}_{i},\upsilon^{*}_{i}\ge0, i=1, \cdots, N,\nonumber
\end{eqnarray}
where
\begin{align}\label{eqn:r19}
f_{i}(p_{1}^{*},\cdots,p_{n}^{*},\alpha^*)=&\frac{\alpha^* h_{i}}{(\sigma^{2}+\alpha^*\sum_{j=1}^{N}h_{j}p^*_{j})(\sum^{N}_{j=1}p^*_{j}+p_c)}\nonumber \\&-\frac{\ln (1+\frac{\alpha^*}{\sigma^2}\sum_{j=1}^{N}h_{j}p^*_{j})}{(\sum^{N}_{j=1}p^*_{j}+p_c)^2}\nonumber\\&+\mu^*\zeta h_{i}(1-\alpha^*),
\end{align}
%\
\begin{align}\label{eqn:r67}
T(p_{1}^{*},\cdots,p_{n}^{*},\alpha^*)=\frac{1}{(\sum^{N}_{i=1}p^*_{i}+p_c)(\sigma^{2}+\alpha^*\sum_{i=1}^{N}h_{i}p^*_{i})}.
\end{align}
Then we rearrange $f_{i}$ as
\begin{align}\label{eqn:r118}
f_{i}=&h_{i}\biggl(\frac{\alpha^*}{(\sigma^{2}+\alpha^*\sum_{j=1}^{N}h_{j}p^*_{j})(\sum^{N}_{j=1}p^*_{j}+p_c)}\nonumber\\&+\mu^*\zeta (1-\alpha^*)\biggr)-\frac{\ln (1+\frac{\alpha^*}{\sigma^2}\sum_{j=1}^{N}h_{j}p^*_{j})}{(\sum^{N}_{j=1}p^*_{j}+p_c)^2}.
\end{align}

Note that in the right hand side of the equation \eqref{eqn:r118}, the coefficient of $h_{i}$ and the second term are constants and the same for each DA port. Without loss of generality, we sort the channel power gains in descending order, i.e., $h_1>h_2>\cdots>h_N$. Based on this, it is easy to have
\begin{align}\label{eqn:r20}
f_{1}>f_{2}>\cdots>f_{N}.
\end{align}

Before further derivations, we provide some lemmas to give some useful insights.

\begin{lemma}\label{L1}
With a positive $p_{i}$, $f_{i}$ should be non-negative.
\end{lemma}

\emph{Proof:}
In order to have a positive $p_{i}^{*}$, we need $\lambda_{i}^{*}=0$ from \eqref{eqn:r18}. It means $f_{i}=-\lambda_{i}^{*}+\upsilon_{i}^{*}\ge0$ for all $i$.$\hfill\blacksquare$

\begin{lemma}\label{L2}
For any $i$, the following properties hold:

\begin{itemize}
\item Property 1: If $f_{i}<0$, $p^{*}_{j}$ should be zero for $j>i$.

\item Property 2: If $f_{i}>0$, $p^{*}_{j}$ should be $\bar{P}_i$ for $j\leq i$.

\item Property 3: If $f_{i}=0$, the optimal solution should be obtained as $p^{*}_{j}=\bar{P}_j$ for $j<i$ and $p^*_{l}=0$ for $l>i$.
\end{itemize}
\end{lemma}

\emph{Proof:}
See Appendix A.	$\hfill\blacksquare$

With these lemmas, we can further derive that the optimal value of the transmit power $(p_{1}^{*},\cdots,p_{N}^{*})$ is $(\bar{P}_1,\cdots,\bar{P}_{i-1},p_i,0,\cdots,0)|_{0\le p_{i}\le \bar{P}_i}$. With above analysis, now we can solve the problem optimally via the following proposition.

\begin{proposition}\label{P1}
By defining $A_i=\frac{ h_i}{\sigma^2}$, $B_i=1+\frac{ \sum^{i-1}_{j=1}h_j\bar{P}_i}{\sigma^2}-\frac{\bar{E}}{\zeta\sigma^2}$, and $C_i=p_c+\sum^{i-1}_{j=1}\bar{P}_i$, the optimal transmit power of $i$-th DA port  $p_{i}$  and the optimal PS ratio $\alpha$ can be obtained as
\begin{eqnarray}
&&p^{*}_{i} =\left[\tilde{p}_i\right]^{\bar{P}_i}_{P_{\min,i}},\label{eqn:r22}\\
&&\alpha^{*} =\left[1-\frac{ \bar{E}}{\zeta\sum^N_{i=1} h_ip^*_i}\right]^+, \label{eqn:r66}
\end{eqnarray}
where $P_{\min,i}$ and  $\tilde{p_{i}}$ can be obtained by
\begin{align}\label{eqn:r133}
P_{\min,i}=\biggl[\frac{\bar{E}}{\zeta h_{i}}-\frac{\sum_{j=1}^{i-1} h_{j}\bar{P}_{j}}{h_{i}}\biggr]^{+}
\end{align}
\begin{gather}
\tilde{p}_{i}=\left\{\begin{array}{ll}
\frac{1}{A_i}\left[\exp\{\omega(\frac{A_iC_i-B_i}{e})+1\}-B_i\right]&A_iC_i-B_i\ge -1,\\
P_{\min,i}&
\text{otherwise,}\label{eqn:r21}
\end{array}\right.
\end{gather}
\end{proposition}

\emph{Proof:}
See Appendix B. $\hfill\blacksquare$

\noindent Here we define $[x]^+=\max\{x,0\}$ and $[x]_a^b=\min\{b,\max\{x,a\}\}$. In \eqref{eqn:r21}, $\omega(x)$ is the principal branch of the Lambert $\omega$ function, which is defined as the inverse function of $f(x) = xe^x$ \cite{Corless1996}. It is worthwhile to note that with $\frac{\bar{E}}{\zeta}\leq \sum^{i}_{j=1} h_j\bar{P}_j$, $P_{\min,i} \leq \bar{P}_{i}$ is always satisfied.

Now, we derive an optimal policy which maximizes the EE for the system. Firstly, we solve transmit power $p_{1}$ of DA port 1 with the largest channel gain $h_{1}$. We compute the $P_{\min,1}$. If $P_{\min,1}$ is greater than $\bar{P}_{1}$, $p^*_{1}=\bar{P}_{1}$ comes out and then we need to solve $p_{2}$. Else, we can see that there are three complementary cases according to the slackness condition in \eqref{eqn:r18} as $(p^{*}_{1},\lambda^{*}_{1},\upsilon^{*}_{1})=\{(0,\lambda^{*}_{1},0),(\tilde{p}_{1},0,0)|_{P_{\min,1}\le \tilde{p}_{1}\le \bar{P}_1},(\bar{P}_1,0,\upsilon^{*}_{1})\}$. In the first case $(p^{*}_{1},\lambda^{*}_{1},\upsilon^{*}_{1})=(0,\lambda^{*}_{1},0)$, $p^*_{1}$ will be set as $0$ and according to the Property 1 in Lemma \ref{L2}, $p^*_{i}$ for $i=2, \cdots, N$ must be $0$ as well because of $f_1=-\lambda^*_1+\upsilon_{1}^{*} \le 0$. So the system's EE becomes $0$ in this case. It is contradicted to our assumption and actually it never occurs.

Next, the second case $(p^{*}_{1},\lambda^{*}_{1},\upsilon^{*}_{1})=(\tilde{p}_{1},0,0)$ is taken into consideration. Because in this case $p_1^*$ is obtained by equating $f_1$ to zero, $f_1=-\lambda_1^*+\upsilon^*_1$ is 0 with $P_{\min,1}\le \tilde{p}_{1}\le \bar{P}_1$ and by Property 3 in Lemma \ref{L2}, $p^*_i$ $(i=2, \cdots, N)$ also equals $0$. So $p^*_{1}$ can be obtained by
\begin{align}\label{eqn:r23}
p^*_{1}=\frac{1}{A_1}\left[\exp\{\omega(\frac{A_1C_1-B_1}{e})+1\}-B_1\right].
\end{align}

It should be taken into account that if $\tilde{p}_{1}$ exceeds the maximum $\bar{P}_1$, $p^*_{1}$ is $\bar{P}_1$. But $p^*_{1}=\bar{P}_1$ is the third case we will discuss later. So if $P_{\min,1}\le \tilde{p}_{1}\le \bar{P}_1$, $p^*_{1}$ is set as $\tilde{p}_{1}$ and $p^*_i$ $(i=2, \cdots, N)$ is $0$.

In the third case $(p^{*}_{1},\lambda^{*}_{1},\upsilon^{*}_{1})=(\bar{P}_1, 0, \upsilon^{*}_{1})$, we need to determine the optimal value of $p_{2}^{*}$ according to the values of $\lambda_{2}$ and $\upsilon_{2}$ with $p^*_{1}=\bar{P}_1$, because the transmit power of other DA ports has not been decided yet. Then we compute $P_{\min,2}$ and check if $P_{\min,2}\geq \bar{P}_{2}$ is satisfied. If yes, we set $p^*_{2}$ as $\bar{P}_{2}$. Otherwise, we can obtain the solutions corresponding to the remaining cases through making use of the properties in Lemma \ref{L2}. The feasible solutions with $p_{1}^{*}=\bar{P}_1$ and $P_{\min,2}\le \bar{P}_{2}$ are given as
\begin{align}\label{eqn:r24}
(p_{1}^{*},p_{2}^{*},\cdots,p_{N}^{*})=\{&(\bar{P}_1,0,\cdots,0),\nonumber\\
&(\bar{P}_1,\tilde{p}_{2},\cdots,0)|_{P_{\min,2}\le \tilde{p}_{2}\le \bar{P}_2},\nonumber\\
&(\bar{P}_1,\bar{P}_2,p_{3},\cdots,p_{N})\},
\end{align}
where $p_{i}$ $(i=3,\cdots,N)$ means the undetermined power for the remaining DA ports and $\tilde{p}_{2}$ is written as
\begin{align}\label{eqn:r25}
\tilde{p}_{2}=\frac{1}{A_2}\left[\exp\{\omega(\frac{A_2C_2-B_2}{e})+1\}-B_2\right].
\end{align}

Like what we have discussed above, $p^{*}_{2}$ is divided into three mutually exclusive cases as $(p^{*}_{2},\lambda^{*}_{2},\upsilon^{*}_{2})=\{(0,\lambda^{*}_{2},0),(\tilde{p}_{2}, 0,0)|_{P_{\min,2}\le \tilde{p}_{2}\le \bar{P}_2},(\bar{P}_2, 0,\upsilon^{*}_{2})\}$. Then $p^{*}_{2}$ is solved for given $\tilde{p}_{2}$ and $P_{\min,2}$. For the third case  $p^{*}_{2}=\bar{P}_2$, we need to determine the optimal values of $p_{i}^{*},i=3,\cdots,N$, and the further procedures will be needed to verify the feasibility of the solutions. Thus we repeat the same procedures like above, and then optimize the power allocation for the rest DA ports, with the transmit power level of DA ports solved in the previous procedures. After obtaining the optimal power allocation scheme, the optimal PS ratio $\alpha^*$ comes out in \eqref{eqn:r66}. To summarize, an algorithm which solves Problem (P1) optimally is presented in Algorithm \ref {alg:A1}. The time complexity of Algorithm \ref {alg:A1} is $\mathcal{O}(N^2+2N+1)$ when all DA ports are activated.

\begin{algorithm}[tb]
\caption{Optimal algorithm for Problem (P1)  }\label{alg:A1}
\begin{algorithmic}[1]
\STATE Set the channel gain as $h_{1}>h_{2}>\cdots>h_{N}$.
\STATE Compute $p^{*}_{1}$ using \eqref{eqn:r22}.
\WHILE{$i\leq N$}
\IF {$p^{*}_{i-1}=\bar{P}_{i-1}$}	
\STATE Compute $p_{i}^{*}$ using \eqref{eqn:r22} with the DA ports transmit power obtained in previous iteration.
\ELSE
\STATE Set $p^{*}_{i}=0$.
\ENDIF
\ENDWHILE
\STATE Obtain $\alpha^{*}$ in \eqref{eqn:r66}.
\end{algorithmic}
\end{algorithm}
\section{Extension to Multiple IoT Devices Case}\label{se3}
\subsection{System Model and Problem Formulation}

Here we investigate the information transmission in downlink DAS under the more general scenario with multiple IoT devices, consisting of $K$ devices, $N$ DA ports and a CP. We adopt the general frequency-division multiplexing access (FDMA) mode to support multi-user transmission, so that the multiple IoT devices occupy non-overlapping channels. Note that FDMA is easy to be implemented for multiple access scenario and has low complexity in both algorithms and hardwares. Furthermore, FDMA has been already standardized and applied in narrowband-IoT (NB-IoT) systems (please see \cite{Zayas2017} and references wherein).  Thus in this scenario we assume that the whole spectrum is equally divided into $K$ channels and each channel is assigned to one device, for avoiding interference. The PS ratio $\alpha_k$ is denoted as the portion of received signal power for ID and the rest $1-\alpha_k$ is for EH  at device $k$. Let $h_{i, k}$ and $p_{i, k}$ respectively denote the channel gain and the transmit power from DA port $i$ to device $k$. Similar to the single device case, we assume that the CP knows perfect CSI for central processing. As a result, the achievable rate of device $k$ is given by
\begin{align}\label{eqn:r71}
R_k=\frac{1}{K}\ln\left(1+\frac{\alpha_{k}\sum_{i=1}^N h_{i, k}p_{i, k}}{\sigma^{2}}\right).
\end{align}
Thus the system's EE for the multiple IoT devices case can be denoted as
\begin{align}\label{eqn:r100}
\eta_{2}=\frac{R_{\mathrm{total}}}{P_{\mathrm{total}}}=\frac{\sum^K_{k=1}R_k}{\sum^K_{k=1}\sum^{N}_{i=1}p_{i, k}+p_c}.
\end{align}

Note that each device decodes information on its own channel but harvests energy from all channels. Then the harvested energy at device $k$ can be expressed as
\begin{align}\label{eqn:r73}
E_k=\zeta(1-\alpha_{k})\sum_{i=1}^{N}h_{i,k}\sum_{k'=1}^{K}p_{i,k'}.
\end{align}
In above, $\sum_{k'=1}^{K}p_{i,k'}$ is the total transmit power of DA port $i$ and $h_{i,k}\sum_{k'=1}^{K}p_{i,k'}$ means the received power at device $k$ from DA port $i$. With the objective to maximize the system's EE by varying the transmit power of DA ports and the PS ratios of devices, subjected to the minimum harvested power requirement $\bar{E}_{k}$ for each device and the maximum transmit power $\bar{P}_i$ for each DA port, we formulate the problem as
\begin{eqnarray}\label{eqn:r74}
 {\rm (P2):}\nonumber
 \max_{\{\alpha_k\}, \{p_{i,k}\}}&&\eta_{2}  \\
{\rm s.t.}&&E_k\ge \bar{E}_{k}, k=1, \cdots, K,\label{eqn:r75}\\
  && 0\leq \alpha_k\leq 1, k=1, \cdots, K, \label{eqn:r81}\\
    && \sum_{k=1}^K p_{i,k}\leq \bar{P}_i, i=1, \cdots, N.\label{eqn:r76} \\
    && p_{i,k} \geq 0, i=1, \cdots, N,\nonumber\\ && k=1,\cdots, K.
   \end{eqnarray}

\subsection{Proposed Suboptimal Algorithm}

Since the objective function $\eta_2$ of Problem (P2) and minimum harvested energy requirements \eqref{eqn:r75} are both non-concave over $\{p_{i,k}\}$ and $\{\alpha_{k}\}$. Finding the optimal solution is difficult and thus we propose a suboptimal algorithm alternatively. Obviously, Problem (P2) is a non-linear fractional programming problem, which can be written as the following form \cite{W.Dinkelbach1967}:
\begin{align}\label{eqn:r13}
q^{*}=\max_{S'\in\mathcal{F}} \frac{R_{\mathrm{total}}(S')}{P_{\mathrm{total}}(S')}.
\end{align}
In \eqref{eqn:r13}, $S'$ is a feasible solution and $\mathcal{F}$ is the feasible set. \eqref{eqn:r13} has another equivalent subtractive form that meets
\begin{align}\label{eqn:r14}
T(q^{*})=\max_{S'\in\mathcal{F}} \left\{ R_{\mathrm{total}}(S')-q^{*}P_{\mathrm{total}}(S') \right\}=0.
\end{align}

It is easy to verify the equivalence between \eqref{eqn:r13} and \eqref{eqn:r14}. The Dinkelbach method in \cite{W.Dinkelbach1967} provides an iterative method to obtain $q^{*}$. Specifically, the problem in the subtractive form with a given $q$  is solved firstly, and then $q$  is updated according to \eqref{eqn:r13}. This iterative process continues until $T(q^{*})$ converges to $0$, which means that $q$ converges to an optimal value. In this paper, we apply this method to address Problem (P2).

The Lagrangian function for a given $q$ can be written as
\begin{align} \label{eqn:r77}
&L_2(\{p_{i, k}\},\{\alpha_k\},\{\upsilon_i\},\{\mu_k\})=\nonumber\\&\frac{1}{K}\sum^K_{k=1}\ln\left(1+\frac{\alpha_{k}\sum_{i=1}^N h_{i, k}p_{i, k}}{\sigma^{2}}\right)-q\left(\sum^{K}_{k=1}\sum_{i=1}^Np_{i, k}+p_{c}\right)\nonumber\\&+\sum^N_{i=1}\upsilon_{i}\left(\bar{P}_{i}-\sum_{k'=1}^Kp_{i, k'}\right)\nonumber\\
&+\sum_{k=1}^K\mu_{k}\left[\zeta(1-\alpha_{k})\sum_{i=1}^{N}h_{i, k}\sum_{k'=1}^{K}p_{i, k'}-\bar{E}_{k}\right].
\end{align}
In \eqref{eqn:r77}, $\{\mu_{k}\}$ and $\{\upsilon_{i}\}$  are Lagrangian multipliers associated with the constraints \eqref{eqn:r75} and \eqref{eqn:r76}, respectively. The dual function is defined as
\begin{equation}\label{eqn:r84}
g_2(\{\upsilon_i\},\{\mu_k\})=\max_{\substack{\{ p_{i, k}\geq 0\} \\ \{0\leq \alpha_{k}\leq 1\}}}L_2(\{p_{i, k}\},\{\alpha_k\},\{\upsilon_i\},\{\mu_k\}).
\end{equation}
Then the dual problem is written as
\begin{align}\label{eqn:r101}
 \min_{\{\upsilon_i\},\{\mu_k\}}g_2(\{\upsilon_i\},\{\mu_k\}).
 \end{align}

Now we consider the maximization problem in \eqref{eqn:r84} for given $\{\upsilon_i\}$ and $\{\mu_k\}$. As the rate expression \eqref{eqn:r71} is non-concave, the optimal solution for problem in \eqref{eqn:r84} is difficult to obtain. Here we propose a two-step suboptimal scheme instead. At the first step, for given $\{\alpha_k\}$, we alternatively optimize each $p_{i, k}$ with other fixed $p_{j, k}$, $\forall j\neq i$, which is known as the BCD method \cite{Richtarik2014}. As $L_2$ is concave over $\{p_{i, k}\}$ for given $\{\alpha_{k}\}$, the BCD method can guarantee that  $\{p_{i,k}\}$ converges to the optimal value $\{p_{i, k}^*\}$. At the second step, we optimize $\{\alpha_{k}\}$ with fixed $\{p_{i, k}\}$ which is obtained in the first step.

To solve $\{p_{i, k}^*\}$, we solve the derivation of $L_2$ with respect to $p_{i, k}$ as
\begin{align}\label{eqn:r78}
\frac{\partial L_2}{\partial p_{i, k}}=&\frac{\alpha_{k} h_{i, k}}{K(\sigma^{2}+\alpha_{k } \sum_{i=1}^{N}h_{i, k}p_{i, k})}+D_{i},
\end{align}
where $D_i$ is defined as
\begin{align}\label{eqn:r94}
 D_i=-q-\upsilon_{i}+\sum^K_{k=1}\mu_{k}\zeta(1-\alpha_{k})h_{i, k}.
\end{align}
Note that $D_i$ is a constant for $k$. There are two cases of $\frac{\partial L_2}{\partial p_{i, k}}$. The first case is $D_i\ge 0$, where $\frac{\partial L_2}{\partial p_{i, k}}$ is positive and $L_2$ is increasing with $p_{i, k}$. Thus $p_{i, k}^*$ equals to $\bar{P}_i$ due to the constraint  \eqref{eqn:r76}. The other case is $D_i< 0$, where $p_{i, k}^*$ can be solved through equaling $\frac{\partial L_2}{\partial p_{i, k}}$ to zero under the total transmit power constraint \eqref{eqn:r76} at each DA port. As a result, to maximize $L_2$, we have
\begin{align}\label{eqn:r79}
p_{i, k}^*=\left\{\begin{array}{ll}
\bar{P}_i&D_i\ge 0,\\
\left[-\frac{1}{KD_i}-\frac{\sigma^2}{h_{i, k}\alpha_k}-\frac{\sum^{N}_{j\neq i}h_{j, k}p_{j, k}}{h_{i, k}}\right]_{0}^{\bar{P}_i}&
D_i< 0.
\end{array}\right.
\end{align}

Note again that the BCD optimization of $\{p_{i, k}\}$ by \eqref{eqn:r79} ensures the convergence. Also we can see that in \eqref{eqn:r79}, $p_{i,k}^*$ increases with $h_{i, k}$. This suggests that in order to improve EE, a device with better CSI should be transmitted with higher power since the device is more efficient in wireless power transfer. Next with fixed $\{p_{i, k}\}$, the derivation of $L_2$ with respect to $\alpha_{k}$ is given by
\begin{align}\label{eqn:r78}
\frac{\partial L_2}{\partial \alpha_{k}}=&\frac{\sum^N_{i=1}h_{i, k}p_{i, k}}{K(\sigma^{2}+\alpha_{k} \sum_{i=1}^{N}h_{i, k}p_{i, k})}-\mu_k\zeta\sum^N_{i=1}h_{i, k}\sum^K_{k'=1}p_{i, k'}.
\end{align}
By setting $\frac{\partial L_2}{\partial \alpha_{k}}=0$ under the constraint \eqref{eqn:r81}, the optimal $\alpha_k^*$ can be obtained as
\begin{align}\label{eqn:r82}
\alpha_{k}^*=\left[\frac{1}{K\mu_{k}\zeta\sum^{N}_{i=1}h_{i, k}\sum_{k'=1}^Kp_{i, k'}}-\frac{\sigma^2}{\sum^N_{i=1}h_{i, k}p_{i, k}}\right]^1_0.
\end{align}

Now we obtain $\{\alpha_{k}^*\}$ and $\{p_{i,k}^*\}$, which are the solutions of $g_2(\{\upsilon_i\},\{\mu_k\})$. After obtaining $g_2(\{\upsilon_i\},\{\mu_k\})$ with given $\{\upsilon_i\}$ and $\{\mu_k\}$, the minimization of $g_2(\{\upsilon_i\},\{\mu_k\})$ over $\{\upsilon_i\}$ and $\{\mu_k\}$ can be efficiently solved by the ellipsoid method. The subgradients required for the ellipsoid method are given by
\begin{align}\label{eqn:r85}
d_{n}=\left\{\begin{array}{ll}
\bar{P}_{n}-\sum_{k=1}^K p_{n,k}^*& n= 1,\cdots,N,\\
E_n^*-\bar{E}_{n}&
n= N+1,\cdots,N+K,
\end{array}\right.
\end{align}
where $p_{n,k}^*$ is obtained in \eqref{eqn:r79} and $E_n^*$ is obtained from $\{p_{n,k}^*\}$ and $\{\alpha_k^*\}$. Finally, after obtaining $\{p^*_{n, k}\}$ and $\{\alpha^*_{k}\}$ in the pervious steps, we update $q$ via \eqref{eqn:r13} for next iteration.

Then we solve $\{p^*_{n, k}\}$ and $\{\alpha^*_{k}\}$ again until $q$ converges to an optimal value $q^*$, which is the suboptimal value of the system's EE $\eta_2$. The algorithm for addressing Problem (P2) is summarized in Algorithm \ref {alg:A2}. The time complexity for the BCD method is $\mathcal{O}(KN)$ and the time complexity for the ellipsoid method is $\mathcal{O}((K+N)^{2})$.  Thus the total time complexity for Algorithm 2 is $\mathcal{O}(\kappa (K+N)^{2}KN)$, where $\kappa$ is the number of iterations for updating $q$.

\begin{algorithm}[tb]
\caption{Suboptimal algorithm for solving  Problem (P2)}\label{alg:A2}
\begin{algorithmic}[1]
\STATE Initialize $\{\lambda_i\}\geq 0$ and $\{\mu_k\} \geq 0$, $\{p_{i, k}\}>0$.
\WHILE{$q$ does not converge to a prescribed accuracy}
\WHILE{$\{\lambda_{i}\}$ and $\{\mu_{k}\}$ do not converge}
\STATE Compute $\{p_{i,k}\}$ by \eqref{eqn:r79} using the BCD method.
\STATE Compute $\{\alpha_k\}$ by \eqref{eqn:r82}.
\STATE Update $\{\lambda_i\}$ and $\{\mu_k\}$ using ellipsoid method.
\ENDWHILE
\STATE Update $q$ as \eqref{eqn:r13}.
\ENDWHILE
\end{algorithmic}
\end{algorithm}
By assuming high SNR for Problem (P2), or the noise power $\sigma^{2} \rightarrow 0$, we can write the optimal $\alpha_k$ as
\begin{align}
\alpha_{k}=\min\left(\frac{1}{K\mu_{k}\zeta\sum^{N}_{i=1}h_{i,k}\sum_{k'=1}^Kp_{i,k'}}, 1\right).
\end{align}
For the case $\frac{1}{K\mu_{k}\zeta\sum^{N}_{i=1}h_{i,k}\sum_{k'=1}^Kp_{i,k'}}\leq1$, i.e., $\mu_{k}\zeta h_{i,k}-\frac{h_{i,k}}{K\sum^{N}_{i=1}h_{i,k}\bar{P}_i}\geq 0$, the lower bound of $\alpha_{k}$ is $\frac{1}{K\mu_{k}\zeta\sum^{N}_{i=1}h_{i,k}\bar{P}_i}$.
Based on this, the bound of $D_i$ can be expressed as
\begin{align}
-q-\upsilon_{i} \leq D_i \leq -q-\upsilon_{i}+\sum^K_{k=1}(\mu_{k}\zeta h_{i,k}-\frac{h_{i,k}}{K\sum^{N}_{i=1}h_{i,k}\bar{P}_i}).
\end{align}
And we can rewrite $p_{i, k}^*$ as
\begin{align}\label{eqn:r110}
p_{i,k}^*=\left\{\begin{array}{ll}
\bar{P}_i&D_i\ge 0,\\
\left[-\frac{1}{KD_i}-\frac{\sum^{N}_{j\neq i}h_{j,k}p_{j,k}}{h_{i,k}}\right]_{0}^{\bar{P}_i}&
D_i< 0.
\end{array}\right.
\end{align}
For the other case $\frac{1}{K\mu_{k}\zeta\sum^{N}_{i=1}h_{i,k}\sum_{k'=1}^Kp_{i,k'}} >1$, the upper bound of $\alpha_k$ is $1$ and the value of $D_i$ is $-q-\upsilon_{i}$. Obviously $D_i$ is always negative. Thus $p_{i, k}^*$ can be expressed as
\begin{align}\label{eqn:r9 }
p_{i,k}^*=\left[\frac{1}{q+\upsilon_{i}-K}-\frac{\sum^{N}_{j\neq i}h_{j,k}p_{j,k}}{h_{i,k}}\right]_{0}^{\bar{P}_i}.
\end{align}
Based on above analysis, we can see that the power allocations of the DA ports follow the classical water-filling (WF) solution.

\section{Simulation Results}
\begin{table}[ht]
\renewcommand\arraystretch{1.5}
\caption{\\Simulation Parameters} % title of Table
\centering % used for centering table
\begin{tabular}{|c|c|p{'1'}|} % centered columns (2 columns)
\hline %inserts double horizontal lines
Noise power $\sigma^{2}$ & $-$104 dBm \\  % inserts table heading
\hline % inserts single horizontal line
Path loss exponent & 3 \\
\hline
Length of the square & 10 m \\ %
\hline  % inserting body of the table
Power constraint for the $i$-th DA port & $\bar{P}_i=\bar{P}$\\ %
\hline  % inserting body of the table
Circuit power $p_{c}$  & 0.5 W \\
\hline
DA port deployment & Square layout \\ %
\hline
Energy conversion efficiency $\zeta$ & 0.6 \\ %
\hline
Number of channel realizations & 1000 \\ %
\hline
Number of device generations & 1000 \\ %
\hline %inserts single line
\end{tabular}
\label{table:sim_para} % is used to refer this table in the text
\end{table}
In this section, we evaluate the proposed algorithms for the cases of single and multiple IoT devices via simulation. The main system parameters are shown in Table \ref{table:sim_para}.
\subsection{Single IoT Device Case}
In this subsection, we provide numerical results to evaluate the performance of the proposed optimal solution for the single IoT device case. For comparison purpose, the SE maximization scheme with PS, the conventional suboptimal Dinkelbach scheme (detailed in Appendix \ref{AC}) and the power allocation scheme with a fixed PS ratio $\alpha=0.5$ are considered in simulation. The parameters of the simulation are listed in Table \ref{table:sim_para}. In this DAS, $N$ DA ports are distributed uniformly within a square with an area of  $100$ square meters and the device is randomly distributed within the area. In SE maximization scheme for DAS \cite{Kim2012}, all DA ports transmit signal with full power and the receiver applies PS to meet the minimum harvested energy demand.
\begin{figure}[t]
\begin{centering}
\includegraphics[scale=0.45]{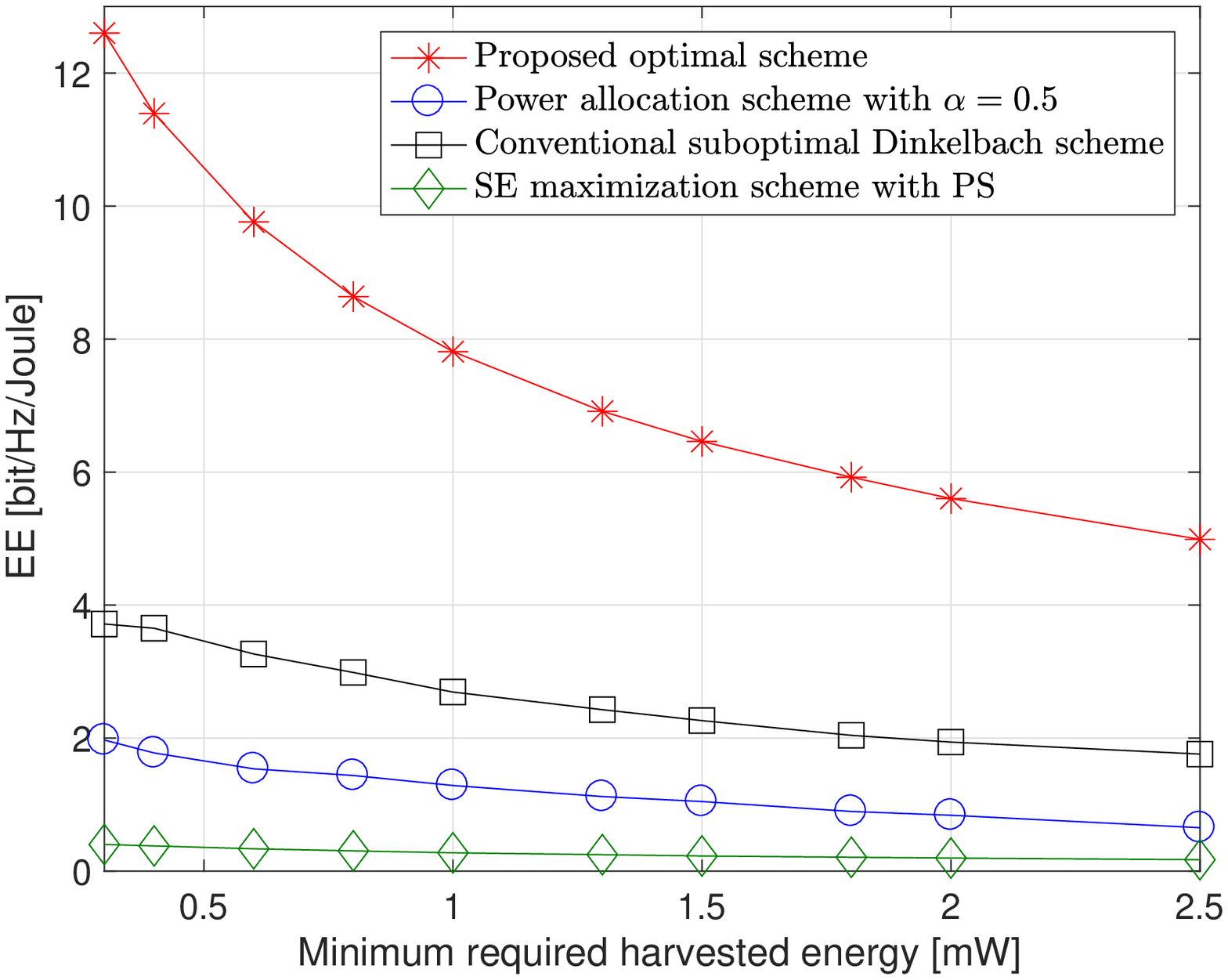}
\vspace{-0.1cm}
 \caption{  Energy efficiency with $N=20$ and $\bar{P}=2 \text{W}$. }\label{fig:simforsig_N5}
\end{centering}
\vspace{-0.1cm}
\end{figure}
\begin{figure}[t]
\begin{centering}
\includegraphics[scale=0.45]{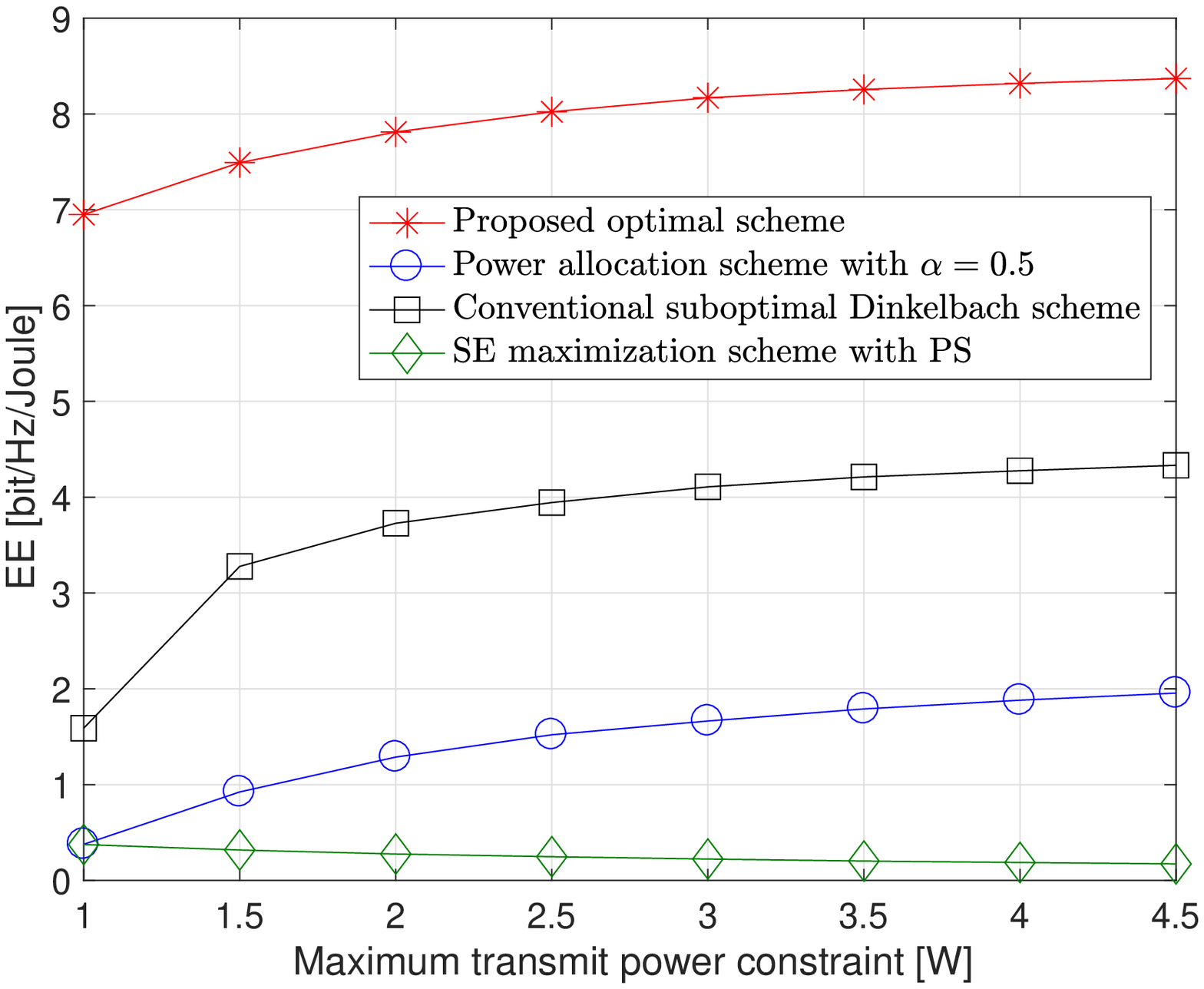}
\vspace{-0.1cm}
 \caption{  Energy efficiency with $N=20$ and $ \bar{E}=1  \text{mW}$.}\label{fig:simmax_N5}
\end{centering}
\vspace{-0.1cm}
\end{figure}
\begin{figure}[t]
\begin{centering}
\includegraphics[scale=0.45]{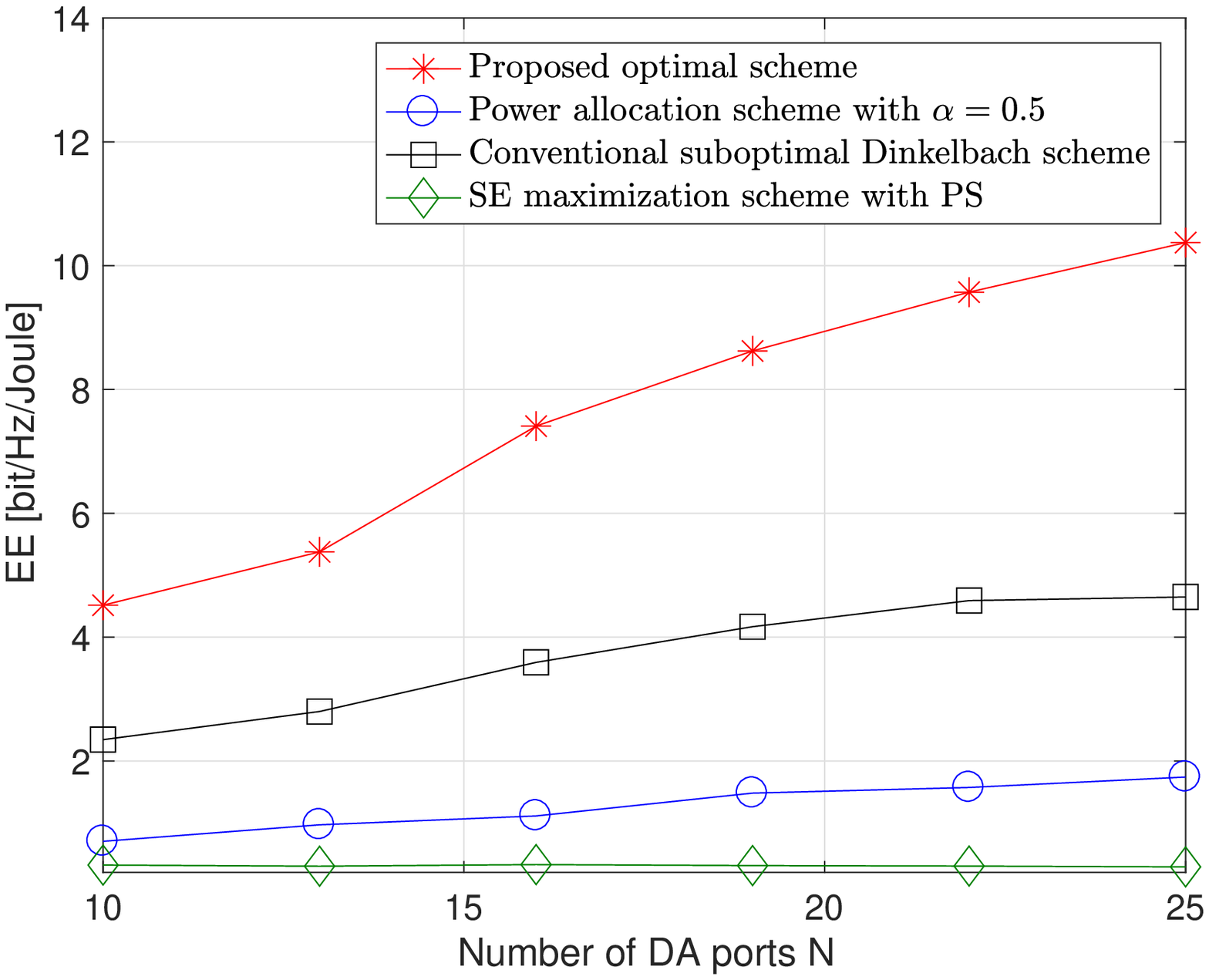}
\vspace{-0.1cm}
 \caption{  Energy efficiency with $\bar{P}=2 \text{W}$ and $ \bar{E}=0.8 \text{mW}$. }\label{fig:sim_N}
\end{centering}
\vspace{-0.1cm}
\end{figure}

In Fig. \ref{fig:simforsig_N5}, a trade-off between EE and the minimum harvested energy requirement $ \bar{E}$ for different schemes is plotted. From Fig. \ref{fig:simforsig_N5}, we observe that our proposed optimal scheme achieves the best performance in terms of EE and provides significant gains over other schemes. As the $ \bar{E}$ increases, the EE performance for all schemes declines due to the reason that the device needs to divide more received power to the energy receiver in order to meet the growing harvested energy requirement. Moreover, the conventional suboptimal Dinkelbach scheme has better EE performance than other benchmark schemes. When it comes to the SE maximization scheme, it has about $96\%$ on average of EE loss and the scheme with a fixed PS ratio $\alpha=0.5$ has about $75\%$ loss in EE on average. But the gap between their EE performance narrows as $ \bar{E}$ increases.

In Fig. \ref{fig:simmax_N5}, we plot the EE performance with  respect to the maximum transmit power constraint $\bar{P}$ where the minimum harvested energy requirement is fixed  as $ \bar{E}=1  \text{mW}$. As the maximum transmit power constraint on each DA port $\bar{P}$ increases, the EE of the proposed optimal scheme, the suboptimal scheme, and the scheme with $\alpha=0.5$ improves and is gradually saturated. As $\bar{P}$ grows, in the proposed optimal scheme, the optimal transmit power in the DA port with best channel condition will be $\tilde{p}_{1}$ in \eqref{eqn:r23} finally rather than $\bar{P}$ and $\tilde{p}_{1}$ will not change anymore, while other DA ports will be turned off, which leads to the saturation of EE in this scheme. The same reason also accounts for the conventional suboptimal Dinkelbach scheme and the power allocation scheme with a fixed PS ratio $\alpha=0.5$. But for the SE maximization scheme where all DA ports transmit with full power, the EE performance in this case drops with the augment of $\bar{P}$, since the denominator of the EE in \eqref{eqn:r1} increases linearly. Hence the gap between the optimal scheme and the SE maximization widens as $\bar{P}$  climbs. As a result, we can demonstrate that our optimal scheme gains the best EE comparing with other benchmark schemes.

In Fig. \ref{fig:sim_N}, we study the relationship between the number of DA ports and the EE of different schemes. When the number of DA ports increases, we can draw the same conclusion that our optimal scheme achieves the best EE performance comparing with other benchmark schemes. There are two benefits of increasing the density of DA ports. The first one is that  the device can harvest more energy from more DA ports. The other is that DA ports are geographically distributed, when there are more DA ports, DA ports within the area become denser and achieve closer access distances between the device and DA ports decline, which results in better EE in the proposed optimal scheme and other benchmark schemes except for the SE maximization scheme. For the SE maximization scheme, the EE performance increases at the first time but becomes worse finally with more DA ports. This is because in this scheme, all DA ports are active and transmit full power, which leads to the linear increase of the denominator in \eqref{eqn:r1}, while the nominator has a logarithmic increase.
\subsection{Multiple IoT Devices Case}
\begin{figure}[t]
%\begin{centering}
\includegraphics[scale=0.45]{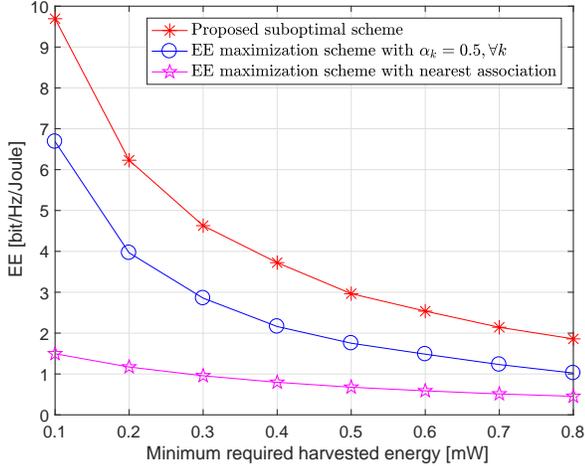}
\vspace{-0.1cm}
 \caption{  Energy efficiency with $\bar{P}=6 \text{W}$ and $4$ devices. }\label{fig:sim_FDM_EE}
%\end{centering}
\vspace{-0.1cm}
\end{figure}
\begin{figure}[t]
%\begin{centering}
\centering
\includegraphics[scale=0.45]{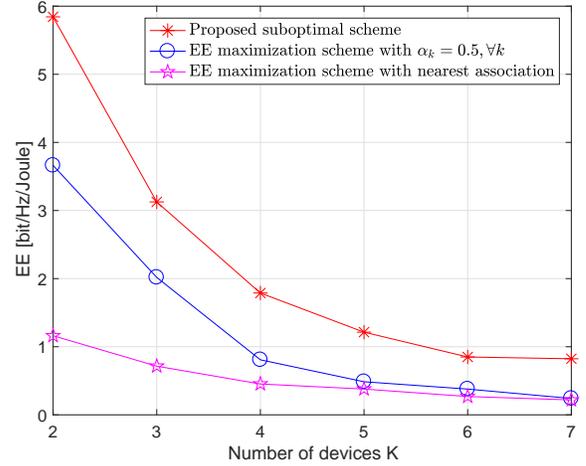}
\vspace{-0.1cm}
 \caption{  Energy efficiency with $\bar{P}=6 \text{W}$ and $\bar{E}=0.8\text{mW}$. }\label{fig:sim_FDM_EE_user}
%\end{centering}
\vspace{-0.1cm}
\end{figure}

Here we evaluate the EE performance of the proposed suboptimal algorithm in Algorithm \ref{alg:A2} through simulations. The system parameters are the same in Table \ref{table:sim_para}. The minimum harvested energy requirement of each device is set as $\bar{E}_{k}=\bar{E}$ for all $k$ in the simulations for convenience. In Fig. \ref{fig:sim_FDM_EE} and Fig. \ref{fig:sim_FDM_EE_user}, we consider the following benchmark schemes: the suboptimal scheme with fixed PS ratios $\alpha_k=0.5, \forall k$ and the EE maximization scheme where each device associates with the nearest DA port but performs EH in all channels (detailed in Appendix \ref{App1}).

 Fig. \ref{fig:sim_FDM_EE} shows the EE versus the minimum required harvested energy $\bar{E}$ by different schemes mentioned above. Firstly, we can confirm that the EE performance of all proposed schemes becomes worse with the increase of minimum required harvested energy $\bar{E}$. The reason accounting for this trend is similar to the single IoT device case and thus it is omitted here. It is worthwhile to note that the gap of EE performance between the proposed scheme and the EE maximization scheme with nearest association narrows when $\bar{E}$ grows. This is due to the fact that in these two schemes, the PS ratio $\alpha_k$ for each device becomes smaller as $\bar{E}_k$ increases and finally approaches to $0$, while the transmit power of each DA port also climbs to $\bar{P}_i$ eventually. Since EE is related to the PS ratio at each device and the transmit power in each DA port, the gap of the EE performance between these two schemes gradually becomes small. For the suboptimal scheme with $\alpha_k=0.5, \forall k$, the gap of EE performance does not narrow as $\bar{E}_k$ increases because the PS ratio $\alpha_k$ in each device is fixed and to satisfy the minimum required harvested energy, the total transmit power $\sum^{K}_{k=1}p_{i, k}$ in each DA port gradually grows to $\bar{P}$. So there must be a performance gap between the proposed scheme and the suboptimal scheme with $\alpha_k=0.5, \forall k$.  Based on the above analysis, we can further derive that in the first place, the EE performance in the scheme with $\alpha_k=0.5, \forall k$ is better than that in the scheme with nearest association because the former scheme has more DA ports to transform information. But the EE in the latter scheme will gradually exceed that in former scheme with $\alpha_k=0.5, \forall k$ as $\bar{E}_k$ ascends.

From Fig. \ref{fig:sim_FDM_EE_user}, we can see that the EE performance among these three schemes declines as the number of devices $K$ increases. This is because  the pre-log factor $1/K$ in \eqref{eqn:r71} is in inverse proportion with the growing number of devices. Moreover, each DA port needs to transmit higher power and each device has to divide more received signals for EH so as to meet the growing demand of overall minimum harvested energy requirement in this system with limited resources, which leads to the decrease of the EE performance.

\section{Conclusions}
In this paper, we studied the problem of EE maximization in SWIPT-based downlink DAS by varying transmit power allocation of DA ports and received power splitting ratios of devices subject to the minimum harvested energy requirement of each device. In the single IoT device case, we obtained the KKT conditions for the problem and derived some useful properties to eliminate numerical complexity of the optimal closed-form solution. For the multiple IoT devices case, we proposed an efficient suboptimal scheme to address the non-convex problem. Simulation results showed that the proposed schemes substantially outperform other benchmark schemes in both single IoT device case and multiple IoT devices case.
\begin{appendices}
\section{Proof of Lemma \ref{L2}}
\begin{itemize}
\item  Property 1: For $f_{i}<0$, $f_{j}=-\lambda_{j}^{*}+\upsilon_{j}^{*}<0$ needs to keep for $j>i$ because of \eqref{eqn:r17} and \eqref{eqn:r20}. So from the slackness condition \eqref{eqn:r18}, we can obtain $\lambda_{j}^{*}>0$ and $p_{i}^{*}=0$.
\item Property 2: For $f_{i}>0$, $f_{j}=-\lambda_{j}^{*}+\upsilon_{j}^{*}>0$ will keep for $j<i$ because of \eqref{eqn:r17} and \eqref{eqn:r20}. So with the slackness condition \eqref{eqn:r18}, we will find that  $\upsilon_{j}^{*}$ must be positive which leads to $p^{*}_{j}=\bar{P}_i$. And $\lambda_j^*$ equals to $0$ because of \eqref{eqn:r18}. In this condition, only $\upsilon_{j}^{*}>0$ and $p_{j}^{*}=\bar{P}_i$ are satisfied.
\item Property 3: For the conditions \eqref{eqn:r20} and $f_{i}=0$, we can draw a conclusion that $f_{j}>0$ for $j<i$ and $f_{l}<0$ for $l>i$. As a result, with two above-mentioned properties, the optimal solution is that $p^{*}_{j}=\bar{P}_i$ for $j<i$ and $p_{l}=0$ for $l>i$.
\end{itemize}
\section{Proof of Proposition \ref{P1}}
Because the minimum harvested energy requirement $E\ge \bar{E}$ and Lemma \ref{L2}, we have $\alpha^{*} \leq 1 - \frac{E}{\zeta(\sum_{j=1}^{i-1}h_j \bar{P}_j + h_i p_i)} \leq 1 - \frac{E}{\zeta\sum_{j=1}^{i}h_j \bar{P}_j} $. If $1 - \frac{E}{\zeta\sum_{j=1}^{i}h_j \bar{P}_j} \leq 0$, i.e., $\frac{E}{\zeta} \geq \sum_{j=1}^{i} h_j \bar{P}_j$, we have $\alpha^{*} \leq 0$. Thus Problem (P1) is no feasible. As a result, in the following, we only consider the case $\frac{E}{\zeta} < \sum_{j=1}^{i} h_j \bar{P}_j$. Since $\eta_1$ is increasing with $\alpha$, we can further derive that the optimal $\alpha^*$ is achieved at $1-\frac{ \bar{E}}{\zeta\sum^N_{i=1} h_ip^*_i}$ due to the minimum harvested energy constraint \eqref{eqn:r113} for (P1). This is because in order to maximize the EE $\eta_1$, we need to allocate received energy for information decoding as much as possible while meeting the minimum harvested energy requirement. Moreover, as we show the optimality conditions in Lemma \ref{L2}, the optimal transmit power $(p_{1}^{*},\cdots,p_{N}^{*})$ is $(\bar{P}_1,\cdots,\bar{P}_{i-1},p_i,0,\cdots,0)|_{0\le p_{i}\le \bar{P}_i}$. This implies that $N-1$ values of transmit power are constants, i.e., either peak power or zero power, and only one value of transmit power needs to be derived. By assuming that $\alpha$ equals to $1-\frac{ \bar{E}}{\zeta\sum^N_{i=1} h_ip_i}$ with $0\leq\alpha\leq 1$ (otherwise there is no feasible solution for this problem and $\alpha$ as well as $\eta_1$ become $0$) and $(p_{1}^{*},\cdots,p_{N}^{*})=(\bar{P}_1,\cdots,\bar{P}_{i-1},p_i,0,\cdots,0)$, we can reformulate Problem (P1) as
\begin{eqnarray}
 \max_{p_{i}}&&\eta_{1}=\frac{\ln\left(1-\frac{\bar{E}}{\zeta\sigma^2}+\frac{\sum_{j=1}^{i-1} h_{j}\bar{P}_{j}+h_ip_i}{\sigma^{2}}\right)}{\sum_{j=1}^{i-1}\bar{P}_{j}+p_i+p_{c}}  \\
{\rm s.t.}
  && 0\leq p_{i}\leq \bar{P}_i.\label{eqn:r112}
 \end{eqnarray}

Furthermore, because the sign of $\eta_{1}$ is mainly determined by the numerator of $\eta_{1}$, to have a positive EE $\eta_{1}$, the minimum value of $p_{i}$ needs to be set as $P_{\min,i}=\biggl[\frac{\bar{E}}{\zeta h_{i}}-\frac{\sum_{j=1}^{i-1} h_{j}\bar{P}_{j}}{h_{i}}\biggr]^{+}$. Thus Problem (P1) can be further simplified as
\begin{eqnarray}
 \max_{p_{i}}&&\eta_{1}=\frac{\ln\left(1-\frac{\bar{E}}{\zeta\sigma^2}+\frac{\sum_{j=1}^{i-1} h_{j}\bar{P}_{j}+h_ip_i}{\sigma^{2}}\right)}{\sum_{j=1}^{i-1}\bar{P}_{j}+p_i+p_{c}}  \\
{\rm s.t.}
  && P_{\min,i}\leq p_{i}\leq \bar{P}_i.\label{eqn:r112}
 \end{eqnarray}

According to the Lemma 3 in \cite{Kim2015}, for the optimization problem
\begin{eqnarray}\label{eqn:r131}
 \max_{x\geq x_{\min}}g(x)=\max_{x\geq x_{\min}} \frac{\ln(ax+b)}{x+c},
 \end{eqnarray}
with $a\geq 0$, $ax_{\min}+b\geq 1$ and $c>0 $, the optimal solution $x^*$ can be obtained as
\begin{gather}
x^*=\left\{\begin{array}{ll}\label{eqn:r117}
\frac{1}{a}\left[\exp\{\omega(\frac{ac-b}{e})+1\}-b\right]_{x_{\min}}&ac-b\ge -1,\\
x_{\min}&
\text{otherwise,}
\end{array}\right.
\end{gather}
where $[x]_{a}$ is $\max(x,a)$. The proof can be found at the Appendix B in \cite{Kim2015}. So we can apply this Lemma to solve the problem in \eqref{eqn:r112} when $\frac{h_{i}}{\sigma^{2}}>0$, $\sum_{j=1}^{i-1}\bar{P}_{j}+p_{c}>0$ and $1-\frac{\bar{E}}{\zeta\sigma^{2}}+\frac{\sum_{j=1}^{i-1} h_{j}\bar{P}_{j}+h_{i}P_{\min,i}}{\sigma^{2}}>1$ are satisfied. Obviously in \eqref{eqn:r112}, $\frac{h_{i}}{\sigma^{2}}$ and $\sum_{j=1}^{i-1}\bar{P}_{j}+p_{c}$ are always positive. Because we need to have a positive $\eta_1$, $1+\frac{\sum_{j=1}^{i-1} h_{j}\bar{P}_{j}-\frac{\bar{E}}{\zeta}+h_{i}P_{\min,i}}{\sigma^{2}}$ is always greater than $1$. As a result, after utilizing the Lemma 3 in \cite{Kim2015}, the optimal values of $p_{i}$ and $\alpha$ come out in \eqref{eqn:r22} and \eqref{eqn:r66}, respectively.

\section{Conventional Suboptimal Dinkelbach Scheme}\label{AC}

At first, we assume that the device works with a fixed PS ratio $\alpha$ firstly. For the reason that Problem (P1) is a fractional programming problem, we can apply the method \eqref{eqn:r14} into solving Problem (P1). Given $\alpha$, Problem (P1) can be rewritten in an equivalent subtractive form:

\begin{eqnarray}\label{eqn:r3}
 \max_{\{p_{i}\}}&&\ln\left(1+\frac{\alpha\sum_{i=1}^N h_{i}p_{i}}{\sigma^{2}}\right)-q\left(\sum_{i=1}^Np_{i}+p_{c}\right) \nonumber \\
{\rm s.t.}&&E\ge \bar{E},\label{eqn:r114}\\
  && 0\leq p_{i}\leq \bar{P}_i,i=1,\cdots,N.\label{eqn:r132}
 \end{eqnarray}

Obviously it is a standard convex problem over $\{p_{i}\}$ and we can apply the Lagrangian dual method to solve it optimally. So the Lagrangian function for this problem with given $q$ and $\alpha$ can be written as
\begin{align}\label{eqn:r60}
L_3(\{p_{i}\},\mu)=&\ln\left(1+\frac{\alpha\sum_{i=1}^N h_{i}p_{i}}{\sigma^{2}}\right)-q\sum_{i=1}^Np_{i}\nonumber\\&-qp_{c}
+\mu\left[\zeta(1-\alpha)\sum_{i=1}^{N}h_{i}p_{i}-\bar{E}\right].
\end{align}
In \eqref{eqn:r60}, $\mu$ is the Lagrangian multiplier associated with the constraint of $E\ge \bar{E}$. And the dual problem in this scheme is defined as
\begin{align}\label{eqn:r108}
	\min_{\mu}\max_{\{0\leq p_{i}\leq \bar{P}_i\}}L_3(\{p_{i}\},\mu).
\end{align}

Since $L_3$ is concave over $\{p_{i}\}$ for given $\alpha$ and $q$, the BCD method can guarantee that $\{p_{i}\}$ converges to the optimal value $\{p_{i}^*\}$. Thus we adopt the BCD method to solve $\{p_{i}^*\}$. We alternatively optimize each $p_{i}$ with other fixed $p_{j}$, $\forall j\neq i$. To find out the optimal value of $p_i$, we compute the derivation of $L_3$ with $p_i$:
\begin{align}\label{eqn:r61}
&\frac{\partial L_3}{\partial p_{i}}=\frac{\alpha h_{i}}{\sigma^{2}+\alpha \sum_{i=1}^{N}h_{i}p_{i}}-q+\mu\zeta(1-\alpha)h_{i}.
\end{align}

It is obvious that for $\mu\zeta(1-\alpha)h_{i}-q\ge 0$, $\frac{\partial L_3}{\partial p_{i}}$ is always positive for a non-negative $p_i$. So the optimal value of $p$ equals to $\bar{P}_i$ in this situation. For $\mu\zeta(1-\alpha)h_{i}-q <0$,  $\frac{\partial L_3}{\partial p_{i}}$ is a non-increasing function for $p_{i}$. And a solution based on the zero-gradient condition can be obtained by equating $\frac{\partial L_3}{\partial p_{i}}$ to zero and applying the transmit power constraint \eqref{eqn:r132}. Generally, the optimal value of $p_i$ can be written as
\begin{align}\label{eqn:r119}
p^*_{i}=\left\{\begin{array}{ll}
\bar{P}_i&\mu\zeta(1-\alpha)h_{i}-q\ge 0,\\
\left[\tilde{x}_{i}\right]^{\bar{P}_i}_0&
\mu\zeta(1-\alpha)h_{i} -q<0,
\end{array}\right.
\end{align}
where
\begin{align}\label{eqn:r63}
\tilde{x}_{i}=\frac{1}{q-\mu\zeta(1-\alpha)h_{i}}-\frac{\sigma^{2}}{\alpha h_{i}}-\frac{\sum^{N}_{j\neq i}h_{j}p_{j}}{h_{i}}.
\end{align}

Next, we solve the Lagrangian multiplier $\mu$ by bisection method and then update $q$ as \eqref{eqn:r13}. After updating $q$, we solve the $\{p^*_{i}\}$ again until $q$ converges to the optimal value $q^*$. Finally, we obtain $\alpha^*$ by exhaustive search. The optimal $\alpha^*$ in this benchmark scheme is expressed as
\begin{align}\label{eqn:r64}
&\alpha^*=\arg\max_{\alpha} \left(q^*\right).
\end{align}

\section{EE Maximization Scheme with Nearest Association}\label{App1}
This scheme is similar to the proposed scheme in Section \ref{se3} except that each device associates with the nearest DA port but performs EH in all channels. Thus the solution is also suboptimal in this benchmark scheme. The channel gain between the nearest DA port and the device is denoted as
\begin{align}\label{eqn:r86}
&\tilde{h}_{i,k}=\arg\max_{i} \left(h_{i,k}\right).
\end{align}
Thus EE in this scheme can be written as
\begin{align}\label{eqn:r103}
\eta_{3}=\frac{\sum_{k=1}^{K}\ln\left(1+\frac{\alpha_{k}\tilde{h}_{i,k}p_{i, k}}{\sigma^{2}}\right)}{K\left(\sum_{k=1}^{K}\sum_{i=1}^Np_{i,k}+p_{c}\right)}.
\end{align}
%s
The harvested energy for device $k$, $E_k$, is similar to \eqref{eqn:r73}. Referring to Problem (P2), the EE maximization problem in this case can be formulated as
\begin{eqnarray}
 {\rm (P3):}\nonumber
 \max_{\{\alpha_k\},\{p_{i,k}\}}&&\eta_{3}  \\
{\rm s.t.}&&E_k\ge \bar{E}_{k}, k=1,\cdots,K,\label{eqn:r88} \\
  && 0\leq \alpha_k\leq 1,  i=1,\cdots,N,\label{eqn:r89}\\
    && \sum_{k=1}^K p_{i,k}\leq \bar{P}_i,i=1,\cdots,N.\label{eqn:r90}\\&&\nonumber
\end{eqnarray}

As we can see, Problem (P3) is a fractional programming problem, we can also apply the Dinkelbach method \eqref{eqn:r14} in order to decompose Problem (P3) like what we do in section \ref{se3}. So the Lagrangian function for Problem (P3) with given $q$ can be written as
\begin{align}\label{eqn:r91}
&L_4(\{p_{i,k}\},\{\alpha_k\},\{\upsilon_i\},\{\mu_k\})=\nonumber\\&\frac{1}{K}\sum^K_{k=1}\ln\left(1+\frac{\alpha_{k}\tilde{h}_{i,k}p_{i, k}}{\sigma^{2}}\right)-q\left(\sum^{K}_{k=1}\sum_{i=1}^Np_{i, k}+p_{c}\right)\nonumber\\&+\sum^N_{i=1}\upsilon_{i}\left(\bar{P}_{i}-\sum_{k'=1}^Kp_{i,k'}\right)\nonumber\\
&+\sum_{k=1}^K\mu_{k}\left[\zeta(1-\alpha_{k})\sum_{i=1}^{N}h_{i, k}\sum_{k'=1}^{K}p_{i, k'}-\bar{E}_{k}\right].
\end{align}
where $\{\upsilon_{i}\}$ and $\{\mu_{k}\}$ are the non-negative dual variables associated with the corresponding constraints of \eqref{eqn:r90} and \eqref{eqn:r88}, respectively. The dual function is then defined as
\begin{align}\label{eqn:r92}
g_4(\{\upsilon_i\},\{\mu_k\})=\max_{\substack{\{0\leq p_{i,k}\leq \bar{P}_{i}\} \\ \{0\leq \alpha_{k}\leq 1\}}}L_4(\{p_{i,k}\},\{\alpha_k\},\{\upsilon_i\},\{\mu_k\}).
\end{align}
As a result, the dual problem is written as $\min_{\{\upsilon_i\},\{\mu_k\}}g_4(\{\upsilon_i\},\{\mu_k\})$. Now we consider the maximization problem in \eqref{eqn:r92} for solving $ g_4(\{\upsilon_i\},\{\mu_k\})$ with given $\{\upsilon_i\}$ and $\{\mu_k\}$. Because \eqref{eqn:r92} is a non-convex problem, the optimal closed-form solution for this problem is computationally difficult to obtain. Similar to Algorithm \ref{alg:A2}, a two-step suboptimal scheme used for addressing this problem is proposed by us. Firstly, for given $\{\alpha_k\}$, we alternatively optimize each $p_{i, k}$. Because $L_4$ is concave for $\{p_{i,k}\}$ with given $\{\alpha_{k}\}$, this step can guarantee the convergence of  solving $\{p_{i,k}^*\}$. At the second step, we optimize $\{\alpha_{k}\}$ with $\{p_{i,k}\}$ obtained previously.

As we discuss above, in the first place, to address the concave function \eqref{eqn:r92} with fixed $\{\alpha_{k}\}$, we have
\begin{align}\label{eqn:r95}
\frac{\partial L_4}{\partial p_{i,k}}=&\frac{\alpha_{k}\tilde{h}_{i,k}}{K(\sigma^{2}+\alpha_{k}\tilde{h}_{i,k}p_{i,k})}+D_{i},
\end{align}
where $D_i$ equals to \eqref{eqn:r94}. Similar to what we have discussed in the multiple IoT devices case, it exists two mutually exclusively complementary cases to solve the optimal value of $p_{i,k}$ based on $\frac{\partial L_4}{\partial p_{i, k}}$. The first one is $D_i\ge 0$, which leads to a positive $\frac{\partial L_4}{\partial p_{i, k}}$  with $p_{i,k}$. In this case, the optimal value of $p_{i,k}$ is $\bar{P}_i$ due to the constraint \eqref{eqn:r90}. The other case is $D_i< 0$ and then the $p_{i, k}$ maximizing $L_4$ is derived by equating $\frac{\partial L_4}{\partial p_{i, k}}$ to zero and considering the transmit power constraint \eqref{eqn:r90} at each DA port. To sum up, we have
\begin{align}\label{eqn:r93}
p_{i,k}^*=\left\{\begin{array}{ll}
\bar{P}_i&D_i\ge 0,\\
\left[-\frac{1}{KD_i}-\frac{\sigma^2}{\tilde{h}_{i,k}\alpha_k}\right]^{\bar{P}_i}_0&
D_i< 0.
\end{array}\right.
\end{align}
The optimization of $\{p_{i,k}\}$ by \eqref{eqn:r93} ensures the convergence. Next with given $\{p_{i,k}\}$, we have
\begin{align}\label{eqn:r96}
\frac{\partial L_4}{\partial \alpha_{k}}=&\frac{\tilde{h}_{i,k}p_{i, k}}{K(\sigma^{2}+\alpha_{k} \tilde{h}_{i,k}p_{i,k})}-\mu_k\zeta\sum^N_{i=1}h_{i,k}\sum^K_{k'=1}p_{i, k'}.
\end{align}
In \eqref{eqn:r96}, $\frac{\partial L_4}{\partial \alpha_{k}}$ is a non-increasing function with $\alpha_k$. Through setting $\frac{\partial L_4}{\partial \alpha_{k}}=0$ under the constraint \eqref{eqn:r89}, we have
\begin{align}\label{eqn:r98}
\alpha_{k}^*=\left[\frac{1}{K\mu_{k}\zeta\sum^{N}_{i=1}h_{i,k}\sum_{k'=1}^Kp_{i,k'}}-\frac{\sigma^2}{\tilde{h}_{i,k}p_{i,k}}\right]^1_0.
\end{align}

Referring to the Lagrangian dual method in \cite{Boyd2004}, after solving $g_4(\{\upsilon_i\},\{\mu_k\})$ with given $\{\upsilon_i\}$, $\{\mu_k\}$, the minimization of $g_4(\{\upsilon_i\},\{\mu_k\})$ over $\{\upsilon_i\}$, $\{\mu_k\}$ can be obtained by the ellipsoid method efficiently. As a result, by defining $E_n^*=\zeta(1-\alpha_{n}^*)\sum_{i=1}^{N}h_{i, n}\sum_{k'=1}^{K}p_{n, k'}^*$, the subgradients of Problem (P3) required for the ellipsoid method is expressed as
\begin{align}\label{eqn:r99}
d_{n}=\left\{\begin{array}{ll}
\bar{P}_{n}-\sum_{k=1}^K p_{n,k}^*& n= 1,\cdots,N,\\
E_n^*-\bar{E}_{k}&
n= N+1,\cdots,N+k.
\end{array}\right.
\end{align}
In \eqref{eqn:r99}, $p_{n,k}^*$ can be obtained in \eqref{eqn:r93} and $E_n^*$ can be solved with $\{p_{n,k}^*\}$ and $\{\alpha_k^*\}$. After obtaining $\{p^*_{n, k}\}$ and $\{\alpha^*_{k}\}$ in the pervious steps, we update $q$ as \eqref{eqn:r13} for next iteration. Eventually, we solve $\{p^*_{n, k}\}$ and $\{\alpha^*_{k}\}$ again until $q$ converges to the optimal value $q^*$.

\end{appendices}

 \begin{footnotesize}
\bibliographystyle{IEEEtran}
 \bibliography{IoT-2916-2017}

\begin{thebibliography}{10}
\providecommand{\url}[1]{#1}
\csname url@samestyle\endcsname
\providecommand{\newblock}{\relax}
\providecommand{\bibinfo}[2]{#2}
\providecommand{\BIBentrySTDinterwordspacing}{\spaceskip=0pt\relax}
\providecommand{\BIBentryALTinterwordstretchfactor}{4}
\providecommand{\BIBentryALTinterwordspacing}{\spaceskip=\fontdimen2\font plus
\BIBentryALTinterwordstretchfactor\fontdimen3\font minus
  \fontdimen4\font\relax}
\providecommand{\BIBforeignlanguage}[2]{{%
\expandafter\ifx\csname l@#1\endcsname\relax
\typeout{** WARNING: IEEEtran.bst: No hyphenation pattern has been}%
\typeout{** loaded for the language `#1'. Using the pattern for}%
\typeout{** the default language instead.}%
\else
\language=\csname l@#1\endcsname
\fi
#2}}
\providecommand{\BIBdecl}{\relax}
\BIBdecl

\bibitem{Huang2014}
J.~Huang, Y.~Meng, X.~Gong, Y.~Liu, and Q.~Duan, ``A novel deployment scheme
  for green internet of things,'' \emph{IEEE Internet of Things Journal},
  vol.~1, no.~2, pp. 196--205, Apr. 2014.

\bibitem{Huang2016}
J.~Huang, Y.~Yin, Y.~Zhao, Q.~Duan, W.~Wang, and S.~Yu, ``A game-theoretic
  resource allocation approach for intercell device-to-device communications in
  cellular networks,'' \emph{IEEE Transactions on Emerging Topics in
  Computing}, vol.~4, no.~4, pp. 475--486, Oct. 2016.

\bibitem{Huang2017}
J.~Huang, Q.~Duan, C.~C. Xing, and H.~Wang, ``Topology control for building a
  large-scale and energy-efficient internet of things,'' \emph{IEEE Wireless
  Communications}, vol.~24, no.~1, pp. 67--73, Feb. 2017.

\bibitem{Huang2016a}
J.~Huang, Y.~Sun, Z.~Xiong, Q.~Duan, Y.~Zhao, X.~Cao, and W.~Wang, ``Modeling
  and analysis on access control for device-to-device communications in
  cellular network: A network-calculus-based approach,'' \emph{IEEE
  Transactions on Vehicular Technology}, vol.~65, no.~3, pp. 1615--1626, Mar.
  2016.

\bibitem{Lee2012}
S.~R. Lee, S.~H. Moon, J.~S. Kim, and I.~Lee, ``Capacity analysis of
  distributed antenna systems in a composite fading channel,'' \emph{IEEE
  Transactions on Wireless Communications}, vol.~11, no.~3, pp. 1076--1086,
  Mar. 2012.

\bibitem{Choi2007}
W.~Choi and J.~G. Andrews, ``Downlink performance and capacity of distributed
  antenna systems in a multicell environment,'' \emph{IEEE Transactions on
  Wireless Communications}, vol.~6, no.~1, pp. 69--73, Jan. 2007.

\bibitem{Hasegawa2003}
R.~Hasegawa, M.~Shirakabe, R.~Esmailzadeh, and M.~Nakagawa, ``Downlink
  performance of a {CDMA} system with distributed base station,'' in
  \emph{Proc. IEEE 58th Vehicular Technology Conf.. VTC 2003-Fall}, Orlando,
  FL, USA, Oct. 2003.

\bibitem{Lee2013}
S.~R. Lee, S.~H. Moon, H.~B. Kong, and I.~Lee, ``Optimal beamforming schemes
  and its capacity behavior for downlink distributed antenna systems,''
  \emph{IEEE Transactions on Wireless Communications}, vol.~12, no.~6, pp.
  2578--2587, Jun. 2013.

\bibitem{Peng2015}
M.~Peng, K.~Zhang, J.~Jiang, J.~Wang, and W.~Wang, ``Energy-efficient resource
  assignment and power allocation in heterogeneous cloud radio access
  networks,'' \emph{IEEE Transactions on Vehicular Technology}, vol.~64,
  no.~11, pp. 5275--5287, Nov. 2015.

\bibitem{He2013}
C.~He, B.~Sheng, P.~Zhu, X.~You, and G.~Y. Li, ``Energy- and
  spectral-efficiency tradeoff for distributed antenna systems with
  proportional fairness,'' \emph{IEEE Journal on Selected Areas in
  Communications}, vol.~31, no.~5, pp. 894--902, May 2013.

\bibitem{Li2016}
X.~Li, X.~Ge, X.~Wang, J.~Cheng, and V.~C.~M. Leung, ``Energy efficiency
  optimization: Joint antenna-subcarrier-power allocation in {OFDM}-{DAS}s,''
  \emph{IEEE Transactions on Wireless Communications}, vol.~15, no.~11, pp.
  7470--7483, Nov. 2016.

\bibitem{Zhang2010}
J.~Zhang and Y.~Wang, ``Energy-efficient uplink transmission in sectorized
  distributed antenna systems,'' in \emph{Proc. IEEE Int. Conf. Communications
  Workshops}, Capetown, South Africa, May 2010, pp. 1--5.

\bibitem{Chen2012}
X.~Chen, X.~Xu, and X.~Tao, ``Energy efficient power allocation in generalized
  distributed antenna system,'' \emph{IEEE Communications Letters}, vol.~16,
  no.~7, pp. 1022--1025, Jul. 2012.

\bibitem{He2014}
C.~He, G.~Y. Li, F.~C. Zheng, and X.~You, ``Energy-efficient resource
  allocation in {OFDM} systems with distributed antennas,'' \emph{IEEE
  Transactions on Vehicular Technology}, vol.~63, no.~3, pp. 1223--1231, Mar.
  2014.

\bibitem{Kim2015}
H.~Kim, S.~R. Lee, C.~Song, K.~J. Lee, and I.~Lee, ``Optimal power allocation
  scheme for energy efficiency maximization in distributed antenna systems,''
  \emph{IEEE Transactions on Communications}, vol.~63, no.~2, pp. 431--440,
  Feb. 2015.

\bibitem{Kim2012}
H.~Kim, S.~R. Lee, K.~J. Lee, and I.~Lee, ``Transmission schemes based on sum
  rate analysis in distributed antenna systems,'' \emph{IEEE Transactions on
  Wireless Communications}, vol.~11, no.~3, pp. 1201--1209, Mar. 2012.

\bibitem{Liu2013}
L.~Liu, R.~Zhang, and K.~C. Chua, ``Wireless information and power transfer: A
  dynamic power splitting approach,'' \emph{IEEE Transactions on
  Communications}, vol.~61, no.~9, pp. 3990--4001, Sep. 2013.

\bibitem{Zhang2013}
R.~Zhang and C.~K. Ho, ``{MIMO} broadcasting for simultaneous wireless
  information and power transfer,'' \emph{IEEE Transactions on Wireless
  Communications}, vol.~12, no.~5, pp. 1989--2001, May 2013.

\bibitem{Jiang2017}
R.~Jiang, K.~Xiong, P.~Fan, Y.~Zhang, and Z.~Zhong, ``Optimal design of {SWIPT}
  systems with multiple heterogeneous users under non-linear energy harvesting
  model,'' \emph{IEEE Access}, vol.~5, pp. 11\,479--11\,489, 2017.

\bibitem{Xiong2017}
K.~Xiong, B.~Wang, and K.~J.~R. Liu, ``Rate-energy region of {SWIPT} for {MIMO}
  broadcasting under nonlinear energy harvesting model,'' \emph{IEEE
  Transactions on Wireless Communications}, vol.~16, no.~8, pp. 5147--5161,
  Aug. 2017.

\bibitem{Mishra2017}
D.~Mishra and S.~De, ``i\textsuperscript{2}res: Integrated information relay
  and energy supply assisted {RF} harvesting communication,'' \emph{IEEE
  Transactions on Communications}, vol.~65, no.~3, pp. 1274--1288, Mar. 2017.

\bibitem{Ng2013}
D.~W.~K. Ng, E.~S. Lo, and R.~Schober, ``Wireless information and power
  transfer: Energy efficiency optimization in {OFDMA} systems,'' \emph{IEEE
  Transactions on Wireless Communications}, vol.~12, no.~12, pp. 6352--6370,
  Dec. 2013.

\bibitem{Liu2016}
Y.~Liu and X.~Wang, ``Information and energy cooperation in {OFDM} relaying:
  Protocols and optimization,'' \emph{IEEE Transactions on Vehicular
  Technology}, vol.~65, no.~7, pp. 5088--5098, Jul. 2016.

\bibitem{Liu2016a}
Y.~Liu, ``Wireless information and power transfer for multirelay-assisted
  cooperative communication,'' \emph{IEEE Communications Letters}, vol.~20,
  no.~4, pp. 784--787, Apr. 2016.

\bibitem{Xiong2015}
K.~Xiong, P.~Fan, C.~Zhang, and K.~B. Letaief, ``Wireless information and
  energy transfer for two-hop non-regenerative {MIMO}-{OFDM} relay networks,''
  \emph{IEEE Journal on Selected Areas in Communications}, vol.~33, no.~8, pp.
  1595--1611, Aug. 2015.

\bibitem{Zhang2016}
M.~Zhang and Y.~Liu, ``Energy harvesting for physical-layer security in {OFDMA}
  networks,'' \emph{IEEE Transactions on Information Forensics and Security},
  vol.~11, no.~1, pp. 154--162, Jan. 2016.

\bibitem{Zhang2016a}
M.~Zhang, Y.~Liu, and R.~Zhang, ``Artificial noise aided secrecy information
  and power transfer in {OFDMA} systems,'' \emph{IEEE Transactions on Wireless
  Communications}, vol.~15, no.~4, pp. 3085--3096, Apr. 2016.

\bibitem{Ng2015}
D.~W.~K. Ng and R.~Schober, ``Secure and green {SWIPT} in distributed antenna
  networks with limited backhaul capacity,'' \emph{IEEE Transactions on
  Wireless Communications}, vol.~14, no.~9, pp. 5082--5097, Sep. 2015.

\bibitem{Yuan2015}
F.~Yuan, S.~Jin, Y.~Huang, K.~k.~Wong, Q.~T. Zhang, and H.~Zhu, ``Joint
  wireless information and energy transfer in massive distributed antenna
  systems,'' \emph{IEEE Communications Magazine}, vol.~53, no.~6, pp. 109--116,
  Jun. 2015.

\bibitem{Chen2014}
X.~Chen, C.~Yuen, and Z.~Zhang, ``Wireless energy and information transfer
  tradeoff for limited-feedback multiantenna systems with energy beamforming,''
  \emph{IEEE Transactions on Vehicular Technology}, vol.~63, no.~1, pp.
  407--412, Jan. 2014.

\bibitem{Yang2014}
G.~Yang, C.~K. Ho, and Y.~L. Guan, ``Dynamic resource allocation for
  multiple-antenna wireless power transfer,'' \emph{IEEE Transactions on Signal
  Processing}, vol.~62, no.~14, pp. 3565--3577, Jul. 2014.

\bibitem{Zhou2015}
X.~Zhou, ``Training-based {SWIPT}: Optimal power splitting at the receiver,''
  \emph{IEEE Transactions on Vehicular Technology}, vol.~64, no.~9, pp.
  4377--4382, Sep. 2015.

\bibitem{Xu2014}
J.~Xu and R.~Zhang, ``Energy beamforming with one-bit feedback,'' \emph{IEEE
  Transactions on Signal Processing}, vol.~62, no.~20, pp. 5370--5381, Oct.
  2014.

\bibitem{Di2017}
X.~Di, K.~Xiong, P.~Fan, and H.~C. Yang, ``Simultaneous wireless information
  and power transfer in cooperative relay networks with rateless codes,''
  \emph{IEEE Transactions on Vehicular Technology}, vol.~66, no.~4, pp.
  2981--2996, Apr. 2017.

\bibitem{Corless1996}
R.~M. Corless, G.~H. Gonnet, D.~E.~G. Hare, D.~J. Jeffrey, and D.~E. Knuth,
  ``On the {LambertW} function,'' \emph{Advances in Computational Mathematics},
  vol.~5, pp. 329--359, 1996.

\bibitem{Zayas2017}
A.~D. Zayas and P.~Merino, ``The {3GPP} {NB-IoT} system architecture for the
  internet of things,'' in \emph{Proc. IEEE Int. Conf. Communications Workshops
  (ICC Workshops)}, May 2017, pp. 277--282.

\bibitem{W.Dinkelbach1967}
W.~Dinkelbach, ``On nonlinear fractional programming,'' \emph{Management
  Science}, vol.~13, no.~9, pp. 492--498, Mar. 1967.

\bibitem{Richtarik2014}
P.~Richtárik and M.~Takáč, ``Iteration complexity of randomized
  block-coordinate descent methods for minimizing a composite function,''
  \emph{Mathematical Programming}, vol. 144, pp. 1--38, Dec. 2011.

\bibitem{Boyd2004}
S.~Boyd and L.~Vandenberghe, \emph{Convex optimization II}.\hskip 1em plus
  0.5em minus 0.4em\relax Cambridge University, 2004.

\end{thebibliography}
\end{footnotesize}

\end{document}